\documentclass{PoS}

\usepackage{amsmath}

\def\l{\left}
\def\r{\right}

\title{QCD matter with a crossover and a first-order phase transition}

\ShortTitle{QCD matter with a crossover and a first-order phase transition}

\author{\speaker{Christopher J. Plumberg}\\
        Lund University\\
        E-mail: \email{christopher.plumberg@gmail.com}}

\author{Thomas Welle\\
        University of Minnesota\\
        E-mail: \email{welle203.@umn.edu}}

\author{Joseph I. Kapusta\\
        University of Minnesota\\
        E-mail: \email{kapusta@umn.edu}}

\abstract{
We present a phenomenological parametrization of the phase diagram of QCD as a function of temperature $T$ and baryochemical potential $\mu$.  The parametrization is constructed by introducing a switching function which controls the nature of the transition between the Hadron-Resonance Gas (HRG) and Quark-Gluon Plasma (QGP) phases of nuclear matter, such that the equation of state (EOS) possesses a rapid crossover at large $T$ and small $\mu$, a critical point placed anywhere along the phase transition line, and a first-order transition at small $T$ and large $\mu$.  This EOS offers a convenient phenomenological tool for assessing the possible effects of the conjectured QCD critical point on heavy-ion observables.
}

\FullConference{Corfu Summer Institute 2018 "School and Workshops on Elementary Particle Physics and Gravity"\\
		(CORFU2018)\\
		31 August - 28 September, 2018\\
		Corfu, Greece}

\begin{document}

%%%%%%%%%%%%%%%%%%%%%%%%%%%%%%%%%%%%%%%%%%%%%%%%
\section{Introduction}
\label{Sec:intro}
%%%%%%%%%%%%%%%%%%%%%%%%%%%%%%%%%%%%%%%%%%%%%%%%
%\vspace*{-2mm}

One of the primary goals of the program dedicated to relativistic heavy-ion collisions is to understand, as completely as possible, the nature of the transition between the Hadron-Resonance Gas (HRG) and Quark-Gluon Plasma (QGP) phases of nuclear matter \cite{Csernai:1992as,Jacobs:2004qv,QM2014,QM2015,QM2017}.  Some of the most important characteristics of this transition can be presented succinctly in terms of the phase diagram of quantum chromodynamics (QCD) as a function of temperature $T$ and baryon chemical potential $\mu$ \cite{Halasz:1998qr,NSAC,Fukushima:2010bq}.  At large $T$ and small $\mu$, the transition is expected on the basis of lattice calculations to be a rapid crossover \cite{Bernard:2004je,Csernai:2006zz}, whereas it is naturally expected that the phase transition along the $\mu$-axis (with $T=0$) is an actual first-order phase transition \cite{Berges:1998rc,Buballa:2003qv}.  Somewhere along the phase transition line, therefore, it is reasonable to expect that the first-order phase transition is converted into a rapid crossover; the point at which this occurs is commonly known as the QCD critical point \cite{Stephanov:1998dy,Fodor:2004nz,Stephanov:2004wx}.

The existence of the QCD critical point has not yet been experimentally confirmed nor predicted with theoretical certainty \cite{Iwasaki:2003de}.  It is vital that phenomenologically motivated studies of heavy-ion data, such as those employing relativistic hydrodynamics, be capable of accounting for the potential influence of such a critical point on experimental observables \cite{Nonaka:2004pg}.  Virtually all such studies rely on a knowledge of the equation of state (EOS) in order to simulate the evolution of a heavy-ion collision in a way which is consistent with the dynamics of QCD \cite{Huovinen:2009yb}.  However, because of the inherent difficulty in theoretically exploring in the QCD phase diagram, particularly at finite $\mu$ \cite{deForcrand:2010ys}, it is essential to have a way of \emph{phenomenologically} characterizing the phase of nuclear matter which represents the anticipated features of the full QCD phase diagram with reasonable accuracy, even if a complete determination of the QCD EOS from first principles is currently unavailable.

One way of phenomenologically modeling phase transitions in the QCD phase diagram is by defining a function, known as a \textit{switching function}, which allows one to construct a simple interpolation between phases in separate regions of the phase diagram.  This approach was adopted in Ref.~\cite{Albright:2014gva}, where the switching function was defined by a simple function of $T$ and $\mu$ which implemented a rapid crossover everywhere along a fixed phase transition line.  The phenomenological parametrization employed by the authors of that paper worked surprisingly well in describing the lattice at both $\mu=0$ and finite $\mu$ \cite{Borsanyi:2010cj,Bazavov:2012jq,Borsanyi:2012cr,Borsanyi:2014ewa}, and was later used by the same authors to describe net baryon fluctuations in the vicinity of such a rapid crossover \cite{Albright:2015uua}.

The present paper improves upon these earlier successes by constructing a switching function $S$, with values ranging between 0 and 1, which contains not only a rapid crossover at large $T$ and small $\mu$, but also a critical point located at $\l( T_c, \mu_c \r)$, and a first-order phase transition for $T<T_c$.  The location of the critical point must be placed along the coexistence curve (with $S=1/2$), but is otherwise a free parameter of the model.  Our switching function $S$ therefore offers a simple and convenient way of studying the effects of a critical point and first-order phase transition, in conjunction with the rapid crossover, without requiring any detailed knowledge of the microscopic dynamics governing a system of nuclear matter near the coexistence curve.

This paper is organized as follows.  In Sec.\ref{Sec:QCDphaseDiagram}, we present the switching function $S$ and discuss briefly how the various free parameters defining its main features may be chosen. 
We then illustrate the form of $S$ for different combinations of free parameters.  We emphasize that $S$ is infinitely differentiable everywhere except at the critical point and along the line of the first-order phase transition, which is necessary to avoid artificially inducing unphysical, higher-order phase transitions into the model.  In Sec.\ref{Sec:FitSF}, we discuss and present the fitting of our switching function to available lattice data at several values of $\mu$, including $\mu=0$.  The results of our fit are presented in Sec.\ref{Sec:Results}.  Sec.\ref{Sec:Conclusions} finally presents some of our conclusions and recommendations for applying our switching function to other models.

%%%%%%%%%%%%%%%%%%%%%%%%%%%%%%%%%%%%%%%%%%%%%%%%
\section{The QCD phase diagram and the switching function}
\label{Sec:QCDphaseDiagram}
\begin{figure}
	\includegraphics[width=\linewidth]{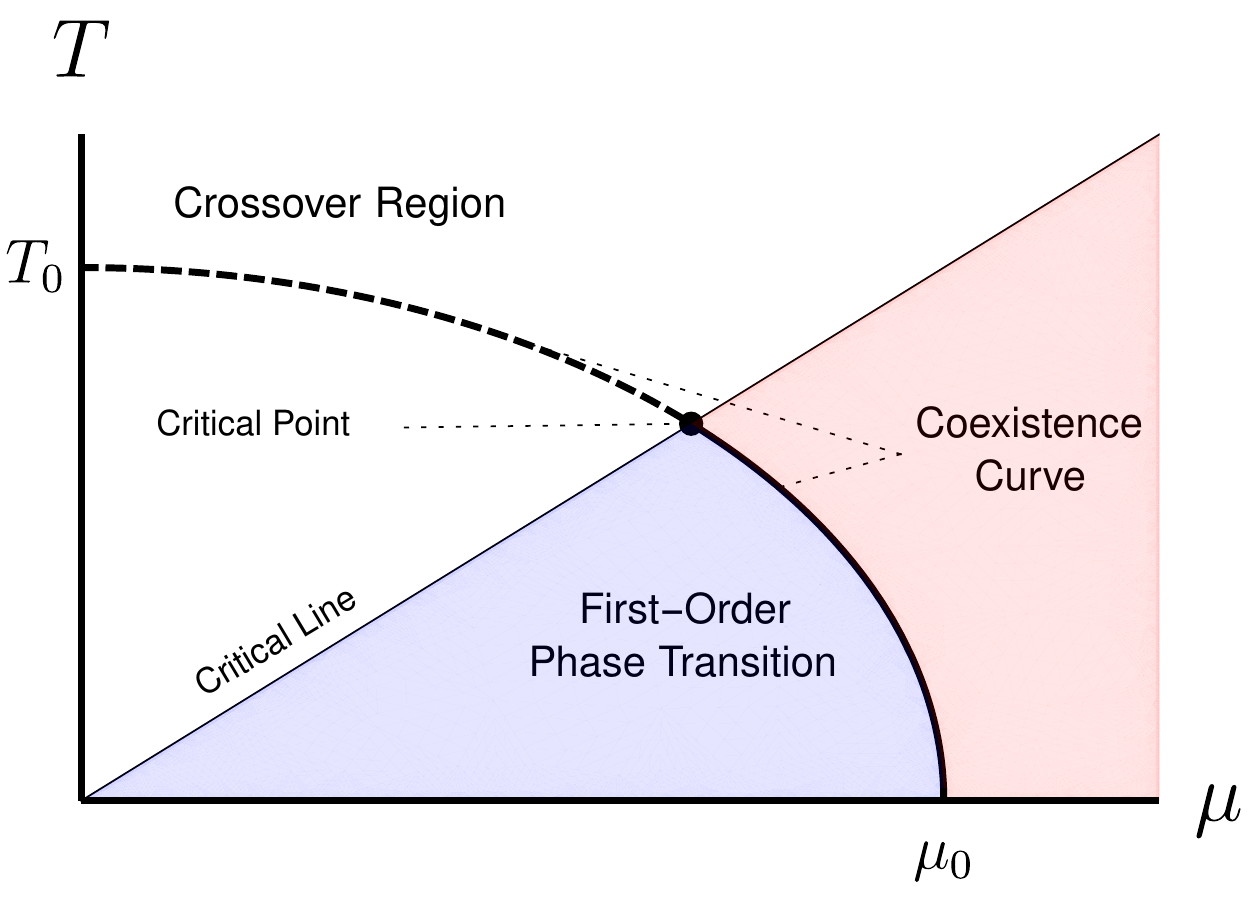}
	\caption{The conjectured structure of the QCD phase diagram, assuming an elliptical coexistence curve.  In this case, the scale parameters $T_0$ and $\mu_0$ are related to the shape and location of the coexistence curve by Eq.~\eqref{coexistenceEllipse}.  The light blue region corresponds to where $S \equiv 0$ (pure hadron-resonance gas), and the light red region shows where $S \equiv 1$ (pure quark-gluon plasma).
	\label{FigQCDPD}}
\end{figure}

%%%%%%%%%%%%%%%%%%%%%%%%%%%%%%%%%%%%%%%%%%%%%%%%
%\vspace*{-2mm}
As we have already discussed, the QCD phase diagram is not completely known.  At large temperatures, the QCD EOS should correspond to the QGP, and at low temperatures, it should correspond to the HRG.  A first-order phase transition is expected to separate the phases at large chemical potential $\mu$, but the transition is actually just a rapid crossover (meaning that thermodynamic quantities like the pressure change rapidly, but all derivatives still remain finite) near $\mu \approx 0$.  The equations of state in each phase can be estimated or computed in various ways, and the idea for this study is to try to model the effects of a first-order phase transition line (and a critical endpoint, where the line switches into the smooth crossover region) by using a switching function to interpolate between the two phases.  If we are interpolating the pressure $P$ in the two phases, this is written as
\begin{equation}
	P(T,\mu) = S(T,\mu) P_{\rm QGP}(T,\mu) + (1-S(T,\mu)) P_{\rm HRG}(T,\mu),
\end{equation}
so that $S=0$ ``turns off" the QGP and ``turns on" the HRG (and \textit{vice versa} for $S=1$).  The crossover/phase transition line then corresponds to $S=1/2$.  To avoid any potential ambiguity, we use the term \textit{coexistence curve} to refer to the contour of the switching function with $S=1/2$; this contour is depicted in Fig.~\ref{FigQCDPD}.  The ray which is located at $\l( T, \mu \r) = \l( 0, 0 \r)$ and passes through the critical point $\l( T, \mu \r) = \l( T_c, \mu_c \r)$, we term the \textit{critical line}.  By construction, $S$ possesses a rapid crossover in the region above the critical line, and a first-order phase transition in the region below the critical line.

As was noted in Ref.~\cite{Albright:2014gva}, discontinuities in the $n$th order derivative correspond to a phase transition of the same order.  $S$ has therefore been constructed to be infinitely differentiable everywhere, except at the critical point and across the first-order phase transition line.

\subsection{The coexistence curve}
\label{SubSec:coexistenceCurve}

The coexistence curve is commonly chosen to be an ellipse, parameterized by
\begin{equation}
		\l(\frac{T}{T_0}\r)^2 + \l(\frac{\mu}{\mu_0}\r)^2 = 1\, , \label{coexistenceEllipse}
\end{equation}
where $T_0$ and $\mu_0$ are scale parameters.  Although this is a convenient choice, we will find the need below to allow the coexistence curve to assume a more general shape, parametrized by a numerically determined curve $R\l( \psi \r)$ in the phase diagram plane, where $\psi \equiv \arctan\l(\mu/T\r)$.  For the particular case of an elliptical coexistence curve, $R(\psi)$ can be written exactly as
\begin{equation}
	R\l(\psi\r)
		= \l[ \l( \frac{\cos \psi}{T_0} \r)^2 + \l( \frac{\sin \psi}{\mu_0} \r)^2 \r]^{-1/2}
	\label{R_psi_coexistence_ellipse}
\end{equation}
In general, the critical point is defined to lie along the coexistence curve at some critical angle $\psi_c$, at a distance from the origin given by $R\l( \psi_c\r)$.

The generalization from an ellipse is especially useful, since the crossover temperature $T_{\mathrm{co}}\l(\mu\r)$ is known from lattice data to be at least a quartic function of $\mu/\mu_c$ \cite{Bazavov:2017dus}:
\begin{equation}
	T_{\mathrm{co}}\l(\mu\r) \equiv T_0\l[ 1 - k_2\l( \frac{\mu}{\mu_c} \r)^2 - k_4\l( \frac{\mu}{\mu_c} \r)^4 - \cdots\r].
	\label{quarticCrossover}
\end{equation}
In this paper, we assume a crossover region with a quartic coexistence curve and disregard any subsequent terms in the $\mu/\mu_c$ expansion.  Furthermore, as we describe below, we constuct $R\l( \psi \r)$ by smoothly joining the quartic curve \eqref{quarticCrossover} and the numerically determined solution to $P_{HRG} = P_{QGP}$ together at the critical point.

\subsection{The switching function}
\label{SubSec:switchingFunction}

In order to retain the excellent agreement between the switching function of Refs.\cite{Albright:2014gva,Albright:2015uua} and the lattice data along the $T$-axis, we require $S$ to reduce to the functional form employed in these earlier works as $\mu \to 0$.  This functional form is given explicitly by
\begin{equation}
	S_{\mathrm{rc}}(T,\mu) = \exp\l[-\theta(T,\mu)\r], \label{Src}
\end{equation}
where
\begin{equation}
	\theta(T,\mu) \equiv \l[ \l( \frac{T}{T_0} \r)^r + \l( \frac{\mu}{\mu_0} \r)^r\r]^{-1}, \label{Albright_theta}
\end{equation}
and the ``rc" underscores that Eqs.~\eqref{Src} and \eqref{Albright_theta} contain only a rapid crossover.  Along the $T$-axis, where $\mu=0$, we have
\begin{equation}
	S_{\mathrm{rc}}(T,\mu=0) = \exp\l[-\l( \frac{T}{T_0} \r)^{-r}\r].	\label{AlbrightEtAl}
\end{equation}
Our new switching function $S$ should have also this form along the $T$ axis.

A function which satisfies the condition \eqref{AlbrightEtAl} and possesses the coexistence curve is given by
%
%\begin{equation*}
%	\mkern-300mu S\l( T, \mu, \psi_c, r \r)
%\end{equation*}
%\vspace{-22pt}
%\begin{equation}
%		= \frac{1}{2} + \frac{1}{\pi}\arg\l( \eta_1\l( m, t, \psi_c \r) + i \eta_2\l( m, t, r \r) \r),
%	\label{SwitchingFunction}
%\end{equation}
\begin{equation}
		S\l( T, \mu, \psi_c, r \r) = \frac{1}{2} + \frac{1}{\pi}\arg\l( \eta_1\l( m, t, \psi_c \r) + i \eta_2\l( m, t, r \r) \r),
	\label{SwitchingFunction}
\end{equation}
where
\begin{eqnarray}
	\eta_1\l( \mu, T, \psi_c \r)
		& \equiv & \frac{1}{2}\l[ 1+ \tanh\l( \frac{a\l( b-\l| \frac{\psi}{\psi_c} \r| \r)}{\l| \frac{\psi}{\psi_c} \r|\l( 1-\l| \frac{\psi}{\psi_c} \r| \r)} \r) \r], \label{eta1}\\
		%\frac{1}{2}\l[ 1+ \tanh\l( \frac{a\l( b-\l| \psi/\psi_c \r| \r)}{\l| \psi/\psi_c \r|\l( 1-\l| \psi/\psi_c \r| \r)} \r) \r], \label{eta1}\\
	\eta_2(\mu, T, r)
		&\equiv & \tan\l[ \frac{\pi}{2^\theta} - \frac{\pi}{2} \r], \label{eta2}\\
		\theta\l( \mu, T, r \r) &=& \l( \frac{T^2+\mu^2}{R^2\l( \psi \r)} \r)^{-r/2}, \label{Rdef}\\
		\psi &=& \arctan\l( \mu/T \r), \label{psidef}
\end{eqnarray}
and
\begin{eqnarray}
	\tan \psi_c &\equiv & \frac{\mu_c}{T_c}. \label{psiCdef}
\end{eqnarray}
We note that $\psi=0$ corresponds to the positive $T$-axis while $\psi=\pi/2$ corresponds to the positive $\mu$-axis.  The critical angle $\psi_c$ divides the phase diagram into two disjoint regions,
\begin{eqnarray}
	&&\l| \psi/\psi_c \r| < 1\text{: crossover region}\nonumber\\
	&&\l| \psi/\psi_c \r| \geq 1\text{: critical region},
\end{eqnarray}
as illustrated in Fig.~\ref{FigQCDPD}.  The function $\arg(z)$ in Eq.~\eqref{SwitchingFunction} is generally defined in terms of elementary functions by
\begin{equation}
	\arg(x+i\,y) \equiv 2 \arctan\l(\frac{y}{x + \sqrt{x^2 + y^2}}\r).
	\label{ArgArcTanIdentity}
\end{equation}
By construction, our switching function is even under $\mu \to -\mu$.  In this paper, we confine our attention to the quadrant of the phase diagram with $T \geq 0$, $\mu \geq 0$.

The most general version of our model contains a total of seven free parameters: $T_c$, $T_0$, $k_2$, $k_4$, $r$, $a$, and $b$.  The parameter $T_c$ fixes the value of the temperature at the critical point.  The next three parameters ($T_0$, $k_2$, $k_4$) are used to determine the curve $R(\psi)$ in the crossover region, as described in more detail below.  The parameter $r$ is a positive integer (typically $\geq 4$) which quantifies the rapidness of the crossover between the two phases in the crossover region, with larger $r$ producing a more rapid crossover transition.  Finally, the parameters $a$ and $b$ control the nature of the transition \emph{along the coexistence curve} from a first-order transition to a simple crossover.  They must satisfy the conditions $a>0$ and $0<b<1$.  Roughly, $a$ quantifies how rapidly the conversion from a first-order transition to a crossover occurs with decreasing $\psi \leq \psi_c$, while $b$ defines approximately where (i.e., at which $\psi$) this conversion happens.  For the results presented in this paper, we have fixed $a=1$ and $b=0.8$.

We evaluate our switching function in the following way.  First, we numerically determine the coexistence curve corresponding to the set of points where the functions $P_{HRG}$ and $P_{QGP}$ are equal to one another.  We then choose a value for the critical temperature $T_c$ and use the point where the coexistence curve crosses this temperature to define the critical chemical potential $\mu_c$.  The critical angle $\psi_c$ is then defined by Eq.~\eqref{psiCdef}.  For angles $\psi \geq \psi_c$, the coexistence curve is defined to coincide with the line of first-order phase transition.  The precise shape of this line cannot be written in a simple algebraic form, and so $R\l(\psi\r)$ in this region must be determined numerically.  Then, below the critical line, the switching function $S$ has the value 0 when $T^2+\mu^2 < R^2(\psi)$ and 1 when $T^2+\mu^2 > R^2(\psi)$.

In the crossover region, we use the quartic curve \eqref{quarticCrossover} to parametrize $R\l(\psi\r)$.  In order to obtain a unique and sufficiently smooth coexistence line, we fix the parameters $T_0$ and $k_2$ appearing in Eq.~\eqref{quarticCrossover} so that both $R\l(\psi\r)$ and its first derivative $R'\l(\psi\r)$ are continuous at the critical point.  The quartic coupling $k_4$ is then retained as one of the free parameters used in optimizing the switching function.  By fixing the shape of $R(\psi)$ in this way, our switching function depends on just three free parameters: $T_c$, $k_4$, and $r$.

%%%%%%%%%%%%%%%%%%%%%%%%%%%%%%%%%%%%%%%%%%%%%%%%
\section{Fitting the switching function}
\label{Sec:FitSF}
%%%%%%%%%%%%%%%%%%%%%%%%%%%%%%%%%%%%%%%%%%%%%%%%
%\vspace*{-2mm}
We now describe the procedure of fitting the switching function to lattice data \cite{Borsanyi:2010cj,Bazavov:2012jq,Borsanyi:2012cr,Borsanyi:2014ewa}.

At this point, our treatment diverges slightly from the work of Refs.~\cite{Albright:2014gva,Albright:2015uua}.  These earlier studies fit to lattice data only along the $T$-axis, and opted to use the corresponding data at finite $\mu$ as a way of demonstrating the surprising versatility of their simple switching function for accurately characterizing the full QCD equation of state (assuming no critical point or first-order transition).  In this paper, we use the available lattice data at both zero and finite $\mu$ to fix an appropriate value of the parameters $r$ and $k_4$, while retaining $T_c$ as a free parameter.  This allows us to construct a switching function which is simultaneously consistent both with currently available lattice data and with placement of a critical point in the (as yet) inaccessible regions of the QCD phase diagram.

\begin{figure}
	\centering
	\includegraphics[width=0.49\linewidth]{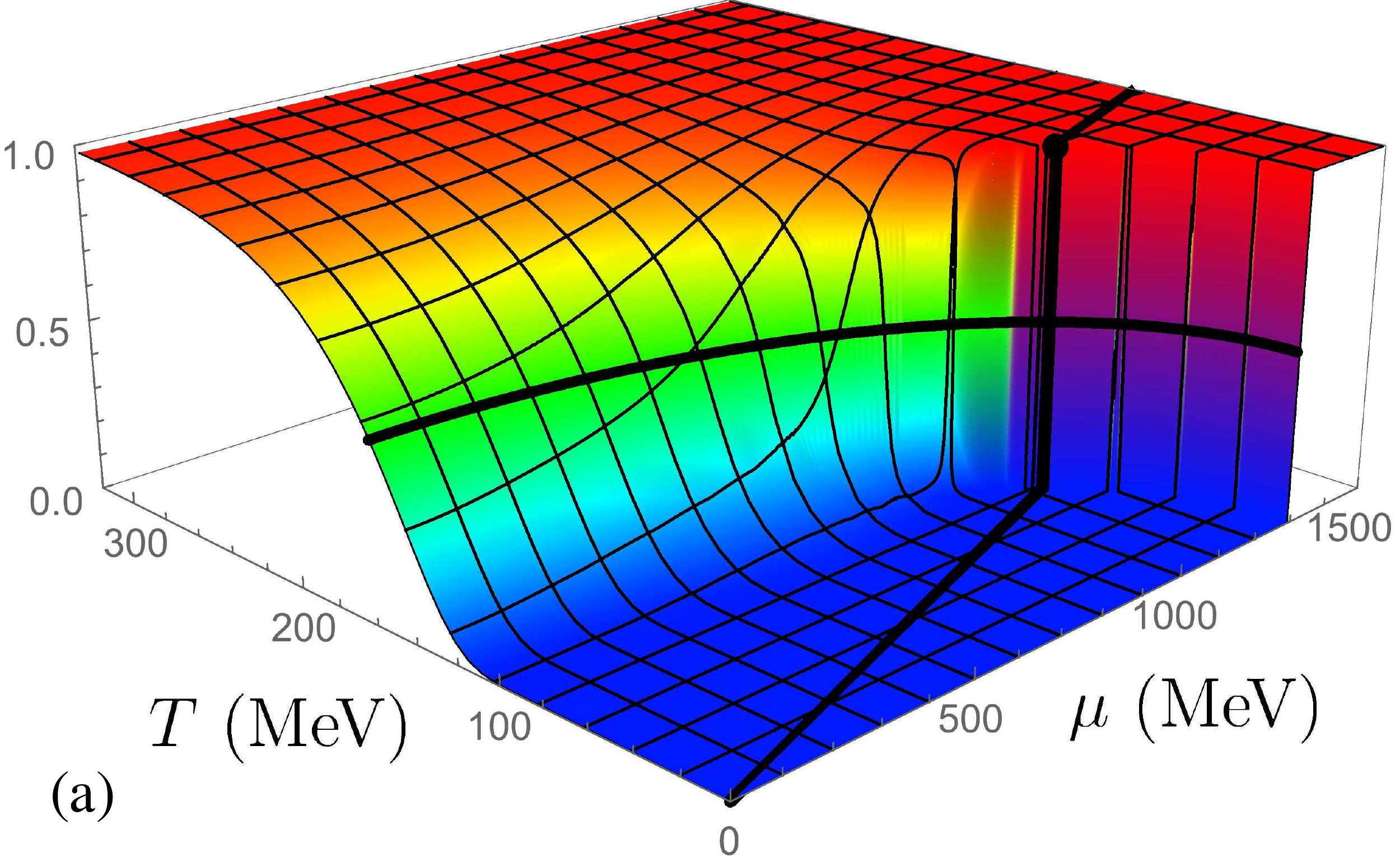}\hfill
	\includegraphics[width=0.49\linewidth]{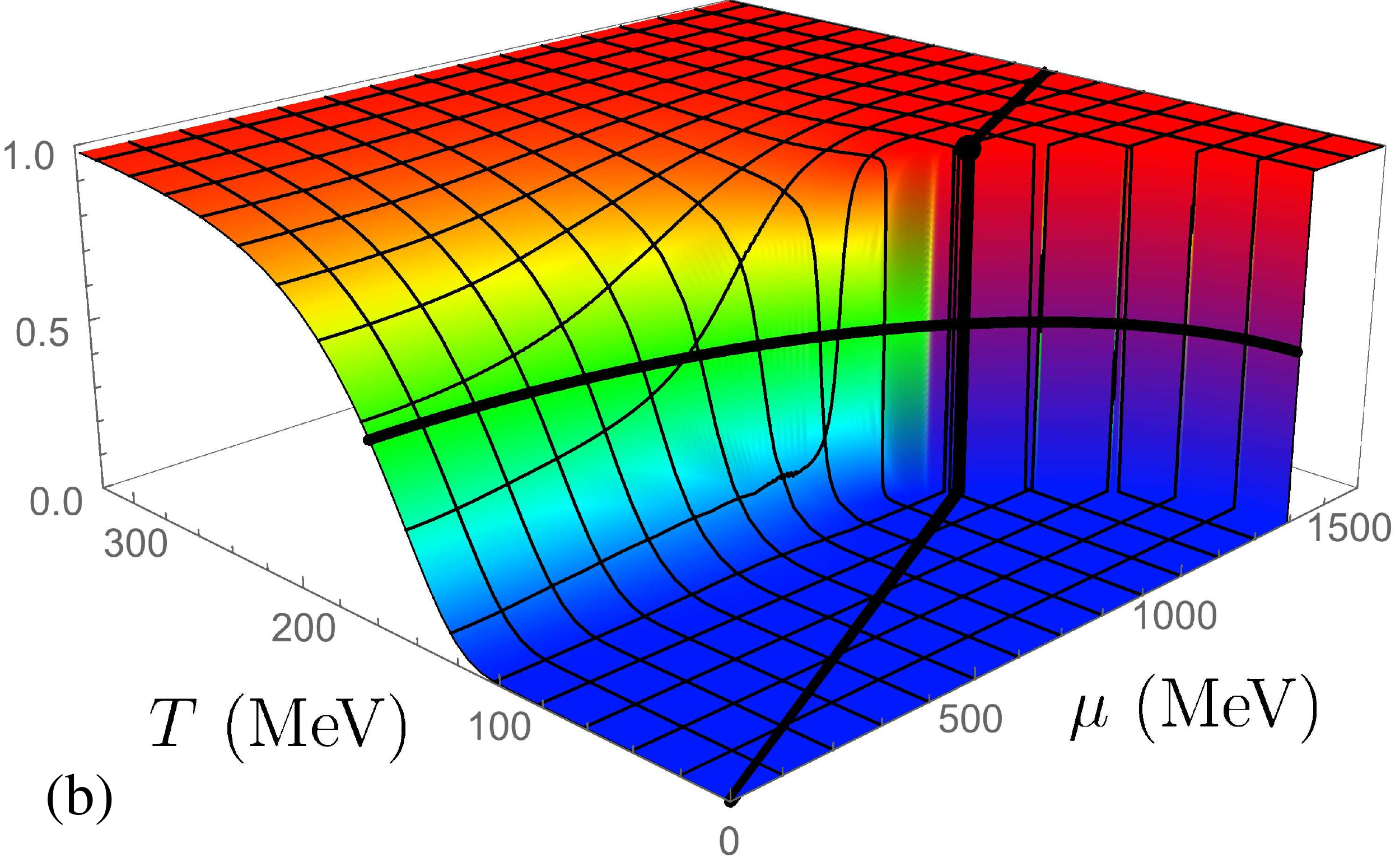}\\
	\vspace{20pt}
	\includegraphics[width=0.49\linewidth]{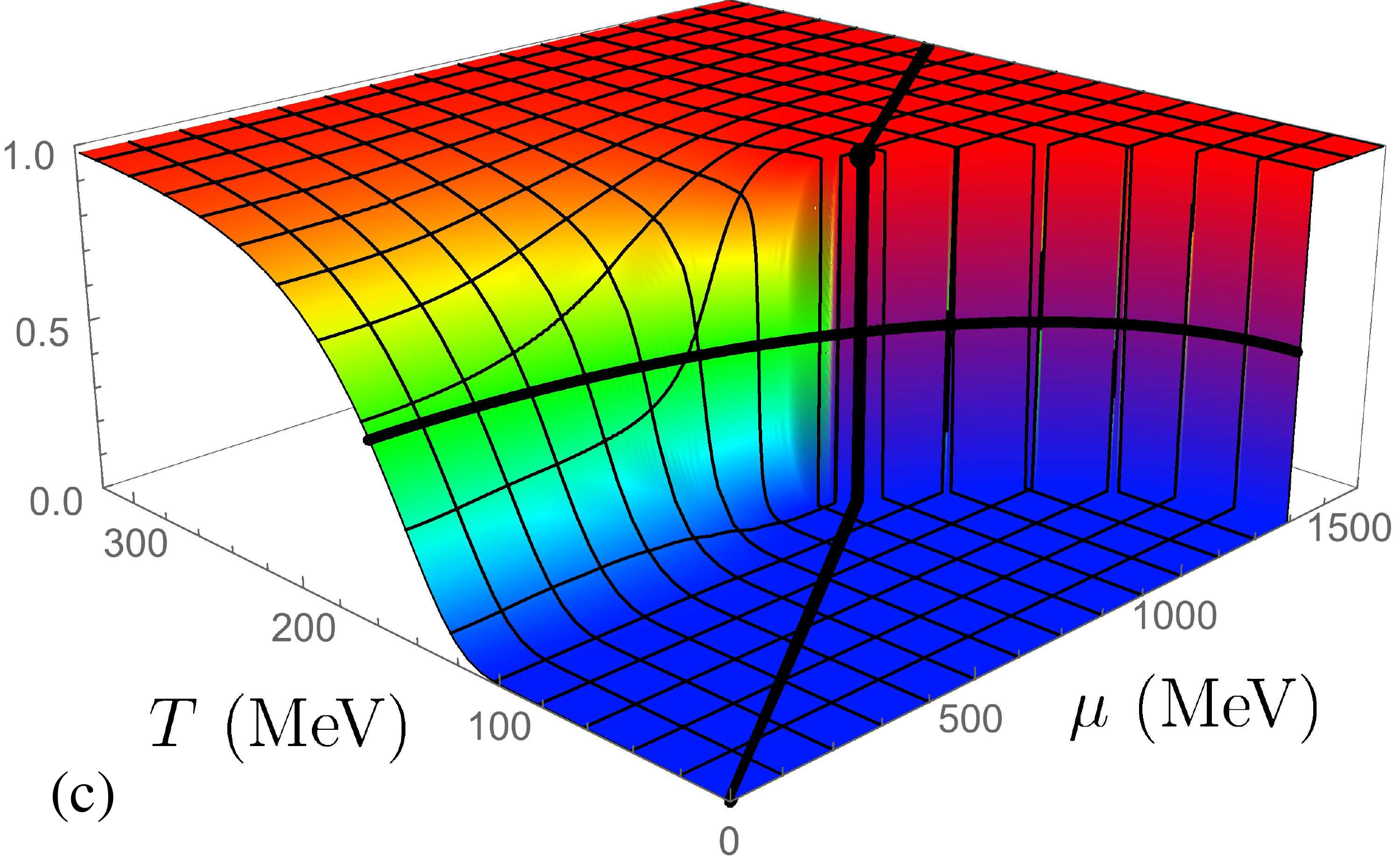}\hfill
    	\begin{minipage}[b]{65mm}% added for caption
    	%\vspace{-50pt}
	\caption{$S$ as given in Eq.~\eqref{SwitchingFunction} with $R\l(\psi\r)$ given by Eq.~\eqref{R_psi_coexistence_ellipse}.  The critical point is shown as a large black point, the entire line $\psi = \psi_c$ is shown as a solid black line, and the thin black line corresponds to the coexistence curve where $S=1/2$.  Panels (a)-(c) correspond to values of $T_c=80,$ $100$ and $120$ MeV, respectively.  Note also that all three plots have precisely the same values along the $T$-axis, as required.
	\label{FigSF}}
	%\vspace{-25pt}
	\end{minipage}
	\vspace{5pt}
\end{figure}

In Fig.~\ref{FigSF}, for illustrative purposes, we plot the form of our switching function for some reasonable choices of the parameters discussed above; these choices do not correspond to the fit results described in the next section.  The specific parameter values used in this Figure are as follows: $T_0 = 150$ MeV, $\mu_0 = 1200$ MeV, $r=5$, $a=1$ and $b=1/2$.  Additionally, we show three different placements of the critical point, specified in terms of the critical temperature $T_c = 80$ MeV (top), 100 MeV (middle), and 120 MeV (bottom).

Several features of the switching function are worth pointing out.  First, note that the function assumes identical values along the $T$-axis, by our requirement of consistency with the lattice data used in Refs.~\cite{Albright:2014gva,Albright:2015uua}.  Second, the properties of the switching function clearly match with intuitive expectations: $S$ is infinitely differentiable everywhere, including in the crossover region, with the sole exception of the critical point itself (large black dot) and the associated first-order line below the critical line (solid black line passing through the origin).  In addition, we have identified the location of the coexistence curve in relation to the switching function, in order to demonstrate that it is indeed a smooth (and, in this case, elliptical) contour corresponding to $S=1/2$ (thick black line tracing coexistence region).

%%%%%%%%%%%%%%%%%%%%%%%%%%%%%%%%%%%%%%%%%%%%%%%%
\section{Results}
\label{Sec:Results}
%%%%%%%%%%%%%%%%%%%%%%%%%%%%%%%%%%%%%%%%%%%%%%%%
%\vspace*{-2mm}

\begin{table}
\begin{center}
    %\begin{tabular}{ | l | l | l | l | l | l | p{5cm} |}
    \begin{tabular}{ | l | l | l | l | l | l | l |}
    \hline
    Model & $T_0$ & $k_2$ & $k_4$ & $r$ & $\chi^2/$d.o.f. \\ \hline
    pt & 197.423 & 0.341508 & 0.000010 & 8 & 9.980 \\ \hline
    exI & 211.8876 & 0.000015 & 0.386452 & 4 & 4.302 \\ \hline
    exII & 212.3557 & 0.000093 & 0.387725 & 4 & 4.574 \\
    \hline
    \end{tabular}
\end{center}
	\caption{The fit results for the three hadronic models considered in this work.  The excluded volume models (exI, exII) clearly offer a superior description of the lattice data than the pt model. \label{TableOfFitResults}}
\end{table}

\begin{figure*}[!htbp]
\minipage{0.49\linewidth}
	\includegraphics[width=\linewidth]{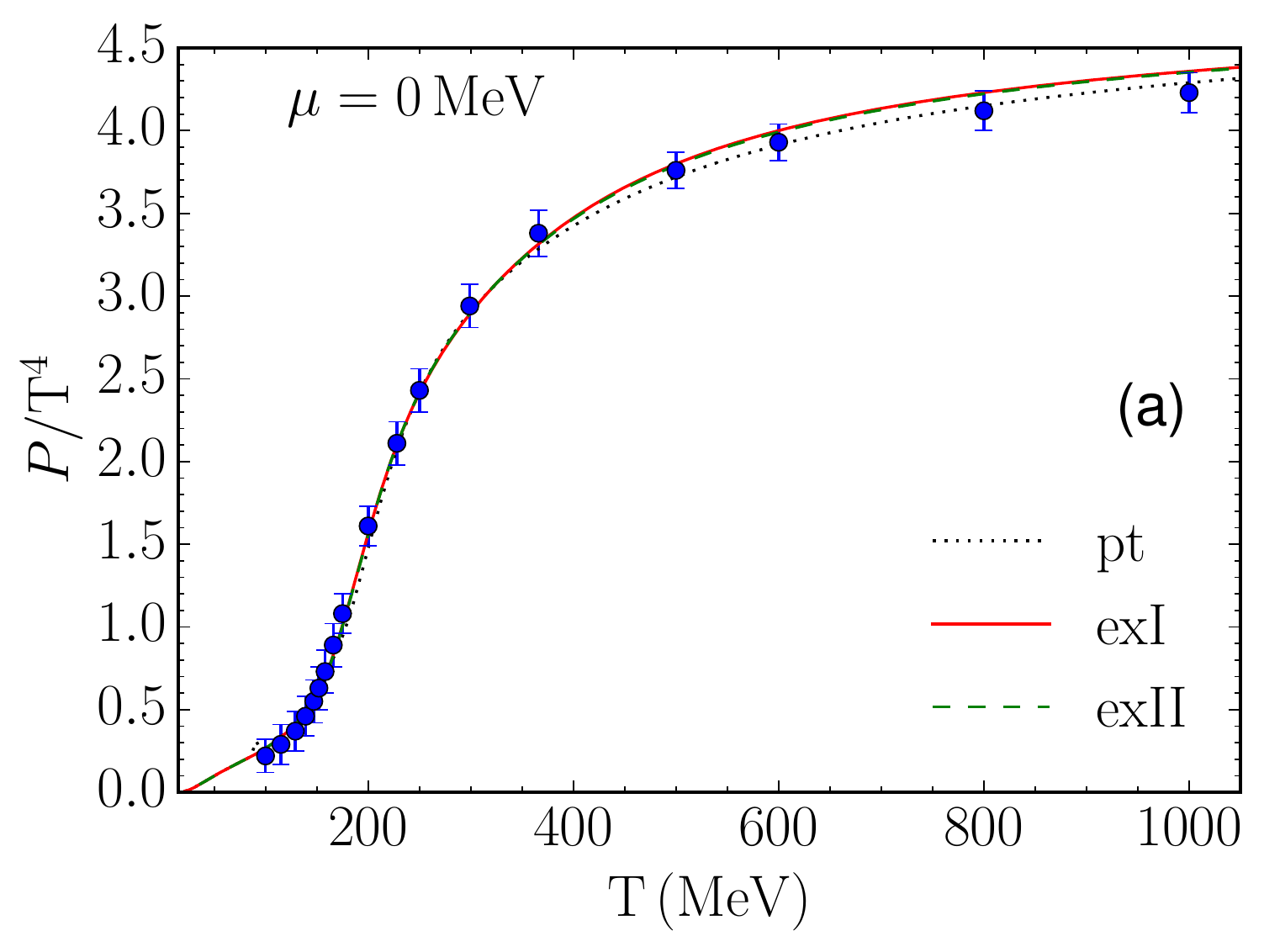}
\endminipage\hfill
\minipage{0.49\linewidth}
	\includegraphics[width=\linewidth]{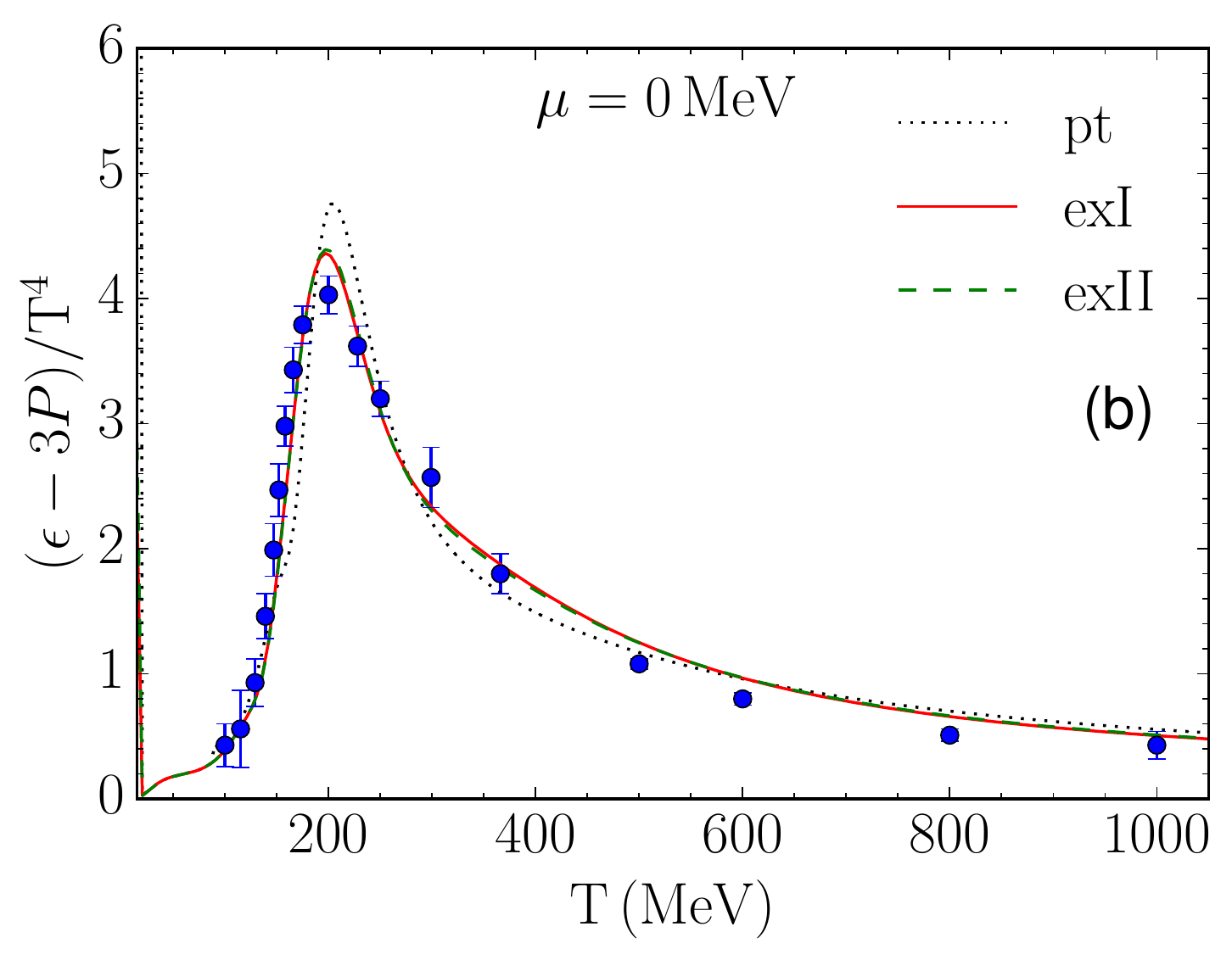}
\endminipage \\
\vspace{10pt}

\minipage{0.49\linewidth}
	\includegraphics[width=\linewidth]{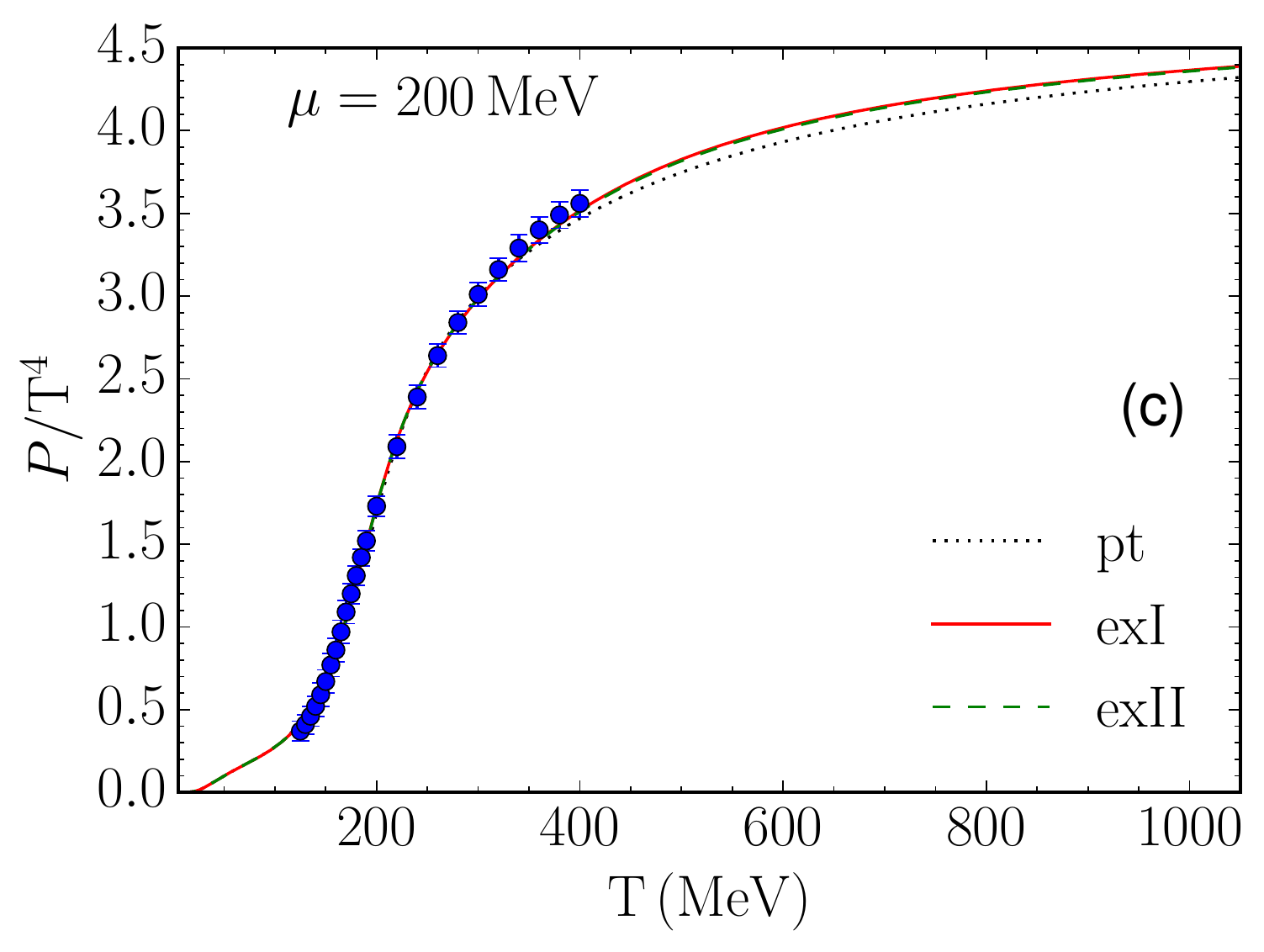}
\endminipage\hfill
\minipage{0.49\linewidth}
	\includegraphics[width=\linewidth]{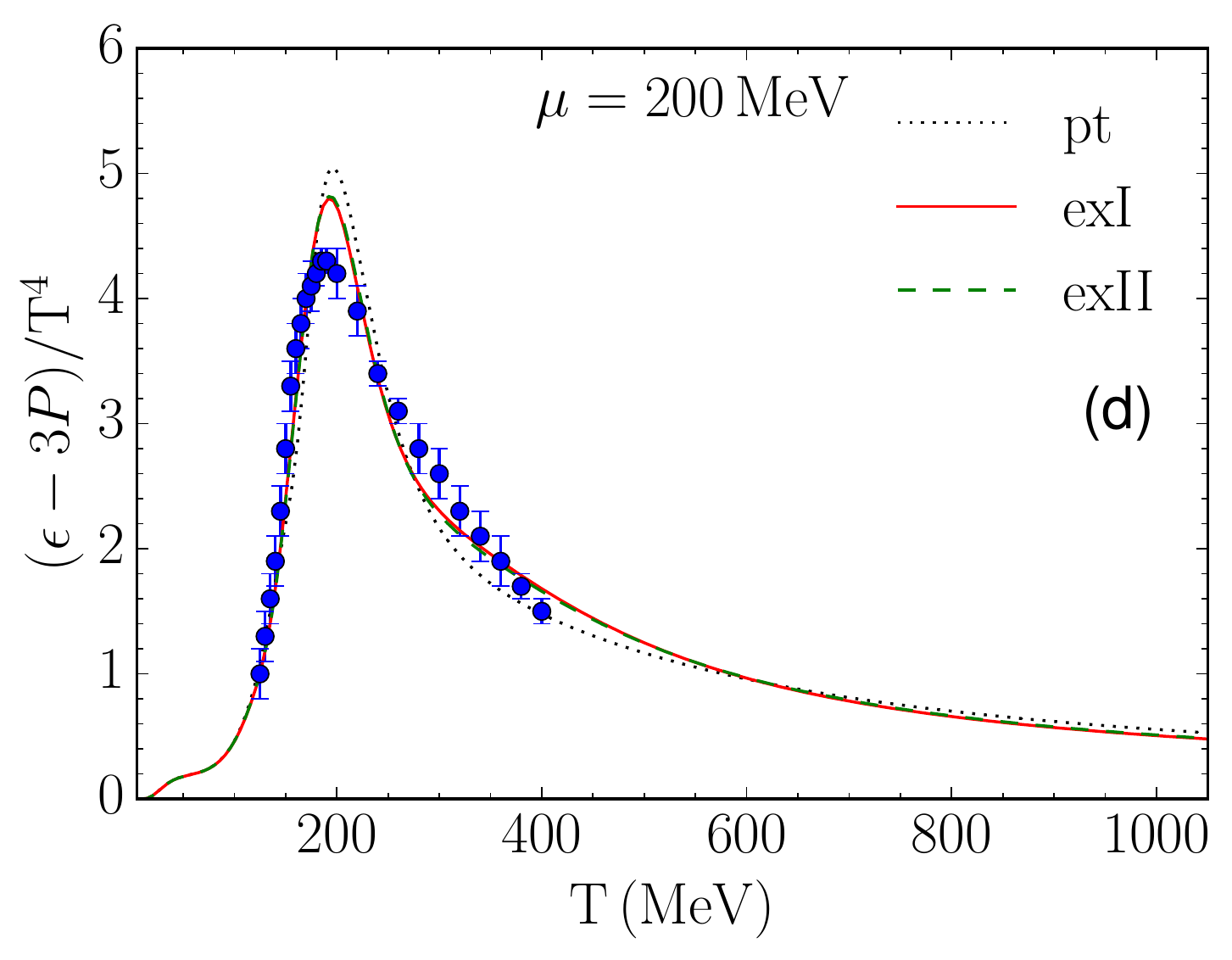}
\endminipage \\
\vspace{10pt}

\minipage{0.49\linewidth}
	\includegraphics[width=\linewidth]{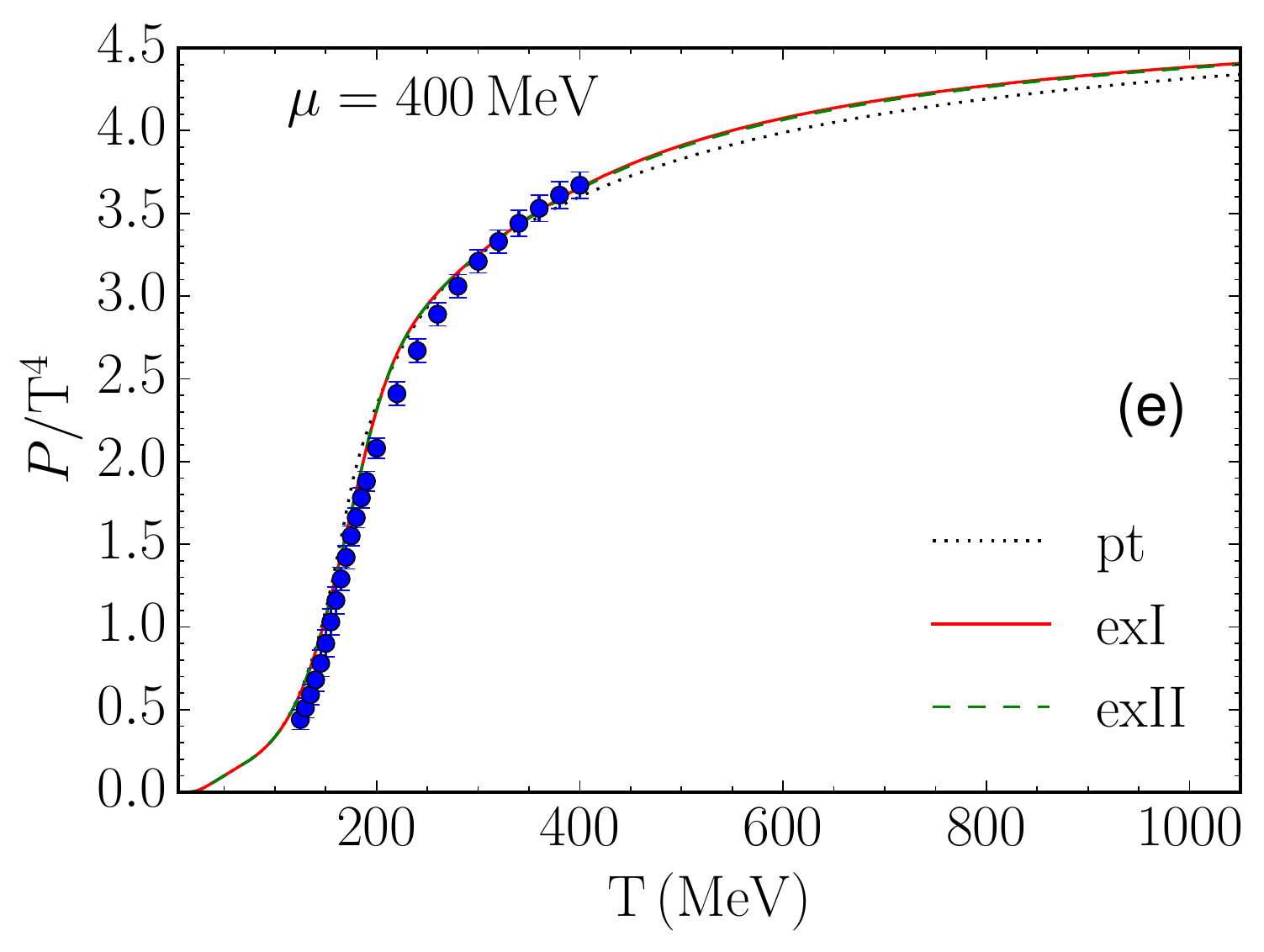}
\endminipage\hfill
\minipage{0.49\linewidth}
	\includegraphics[width=\linewidth]{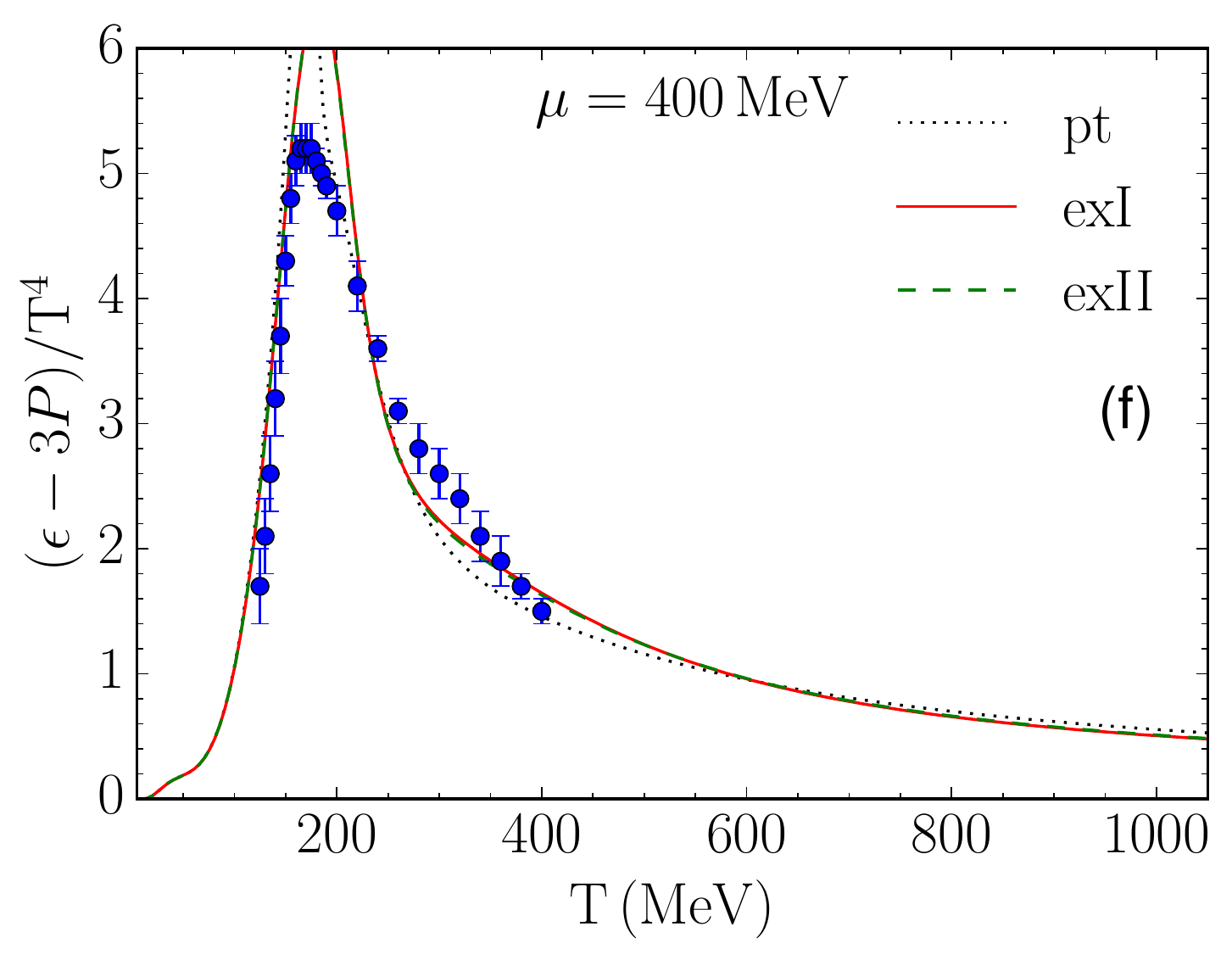}
\endminipage
	\caption{The normalized pressure $P/T^4$ (left panels) and trace anomaly $(\epsilon - 3P)/T^4$ (right panels), for each of the three hadronic models used in the switching function optimization, as compared with lattice calculations taken from \cite{Borsanyi:2012cr}.  The normalized pressure is reproduced reasonably well, but the trace anomaly shows noticeable deviations between the switching function approach and lattice calculations, especially at $\mu \neq 0$.
	\label{FigResultsJointlyOptimized}}
\end{figure*}

\begin{figure}
	\centering
	\includegraphics[width=0.49\linewidth]{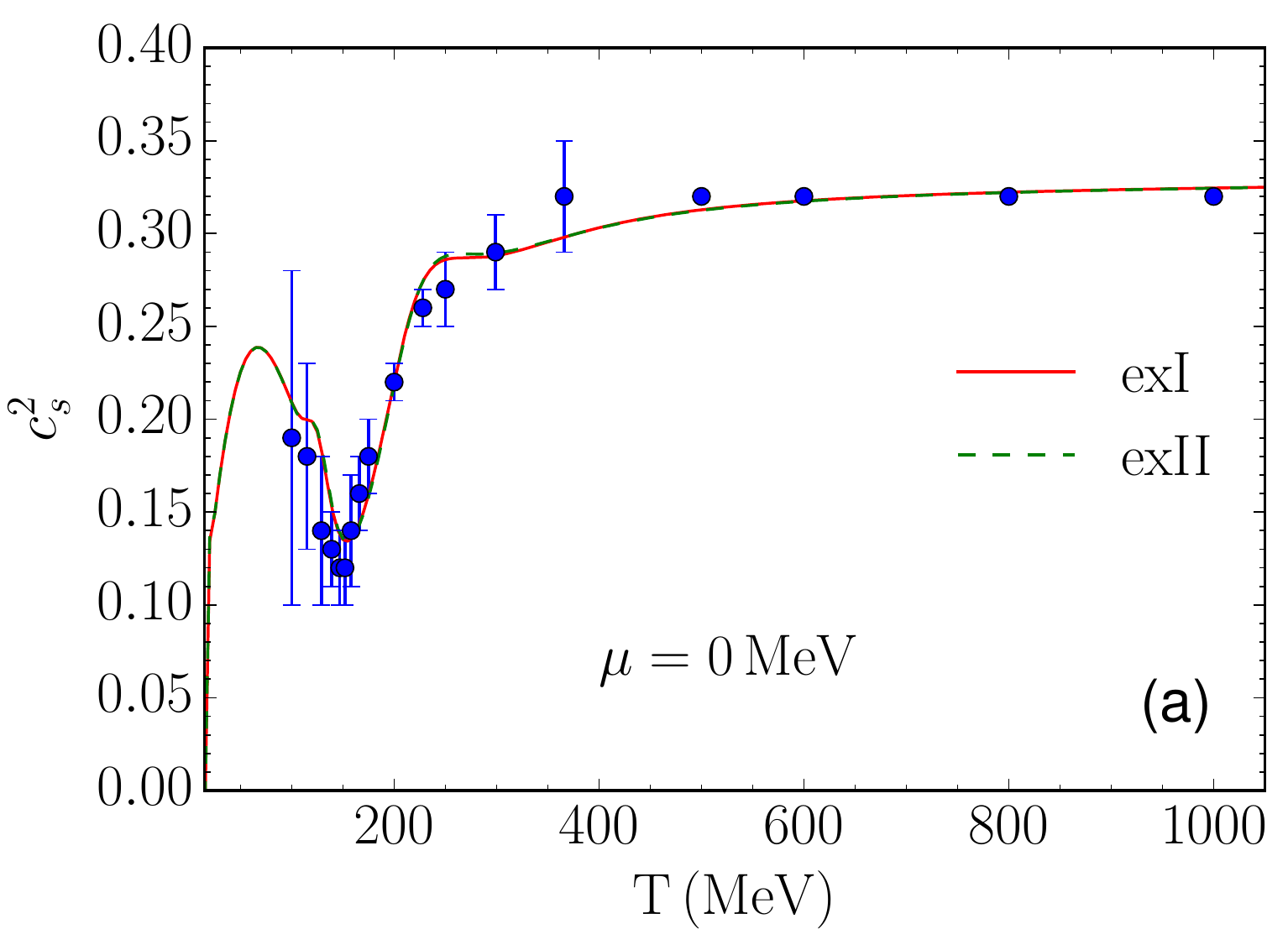}\hfill
	\includegraphics[width=0.49\linewidth]{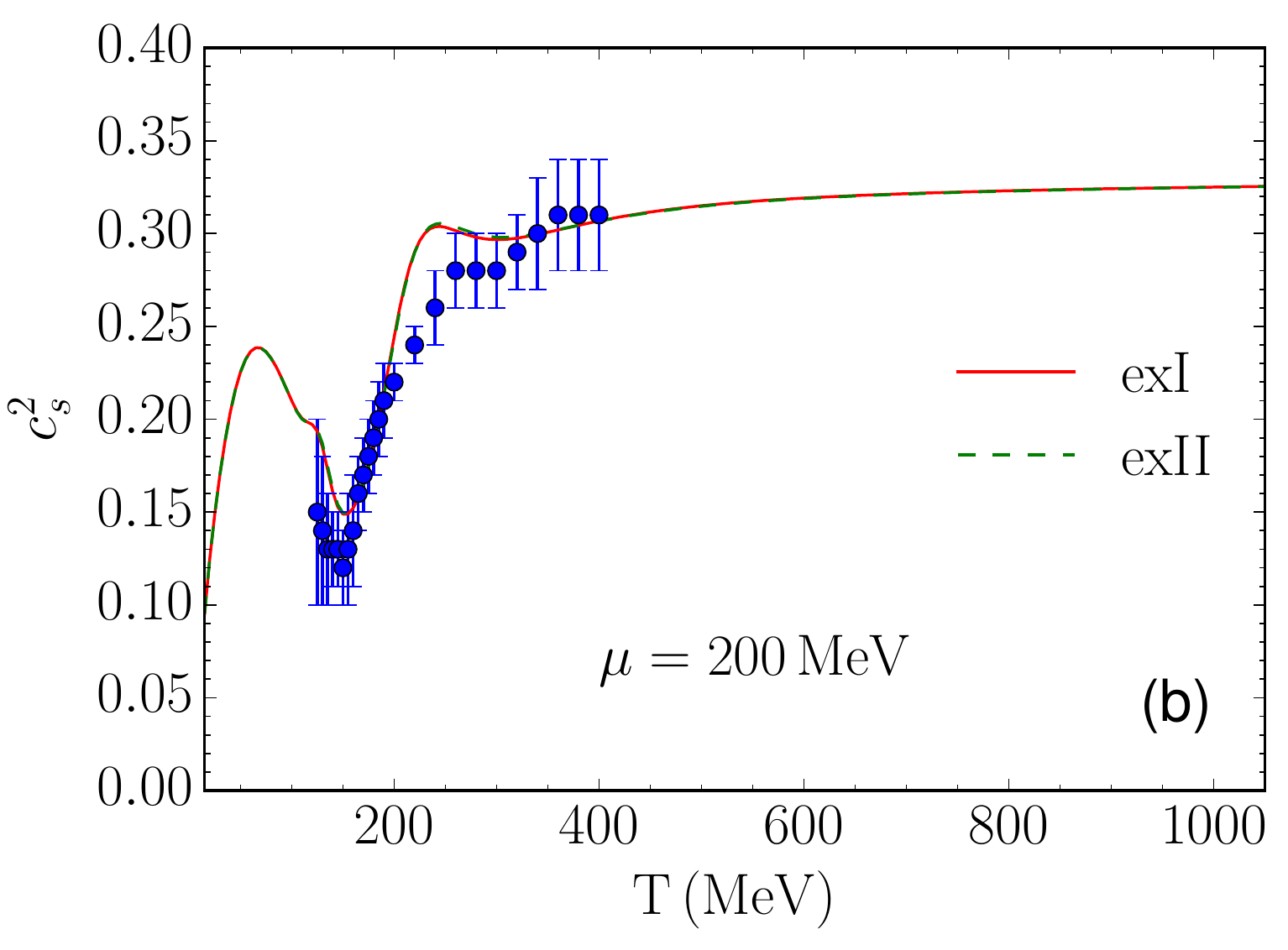}\\
	\vspace{15pt}
	\includegraphics[width=0.49\linewidth]{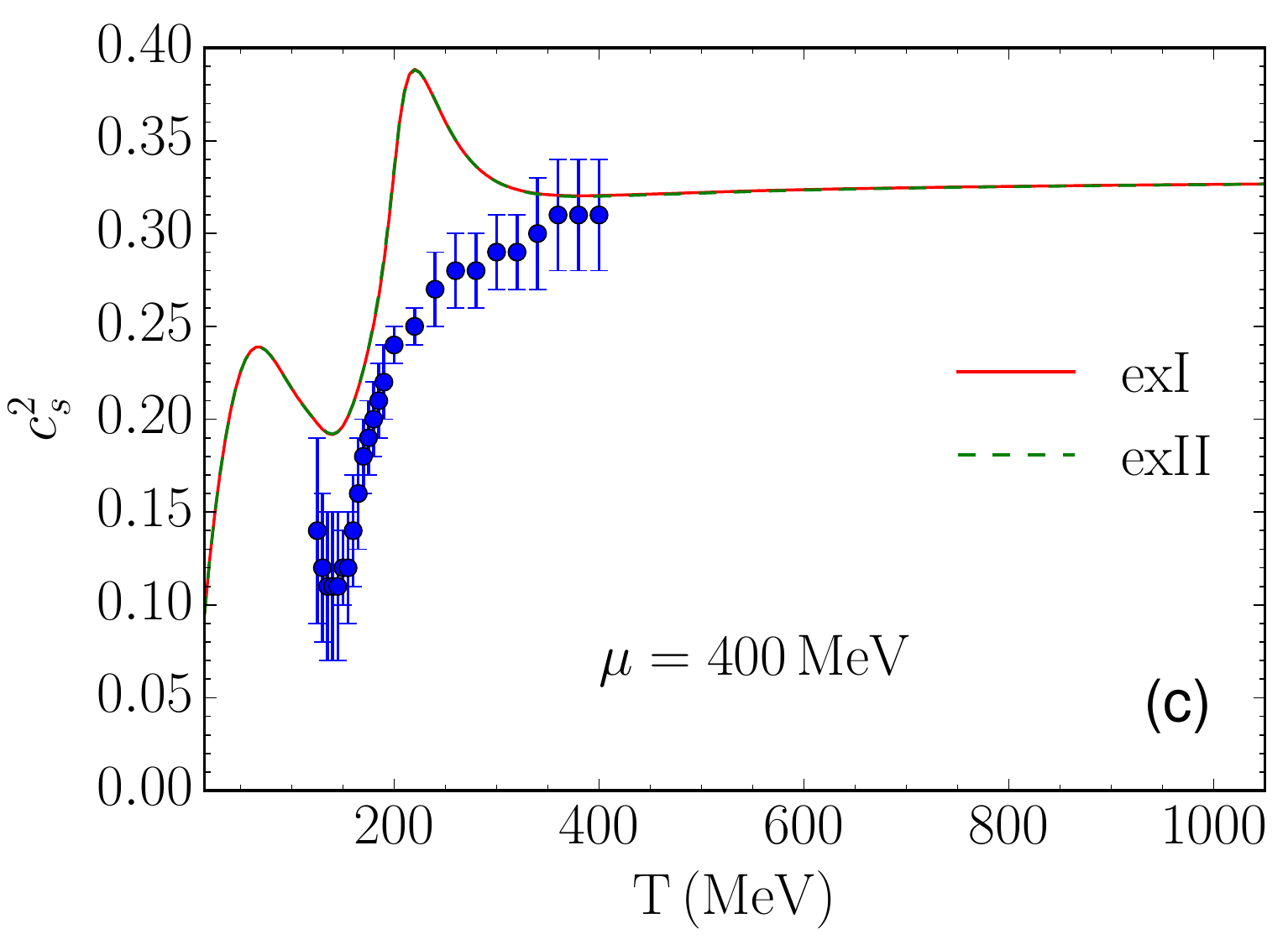}\hfill
    \begin{minipage}[b]{65mm}% added for caption
    %\vspace*{-100pt}
	\caption{Squared speeds of sound $c_s^2$ as functions of temperature $T$, for the two excluded-volume models in the switching function approach: exI (solid red curve) and exII (dashed green curve).  These are compared with lattice calculations (blue points) taken from \cite{Borsanyi:2012cr}.  Agreement with the lattice is apparent at $\mu=0$, but clear discrepancies emerge for $\mu > 0$.
	\vspace{30pt}
	\label{FigSpeedOfSoundResults}}
	\end{minipage}
	\vspace{-10pt}
\end{figure}

Finally, we present the results of fitting our switching function to lattice data, for $T_c=130$ MeV.  We optimize the switching function with respect to two standard lattice observables: the ratio $P/T^4$ and the trace anomaly $\l( \epsilon - 3P \r)/T^2$, as functions of $T$ and $\mu$.  Our fit results are shown in Fig.~\ref{FigResultsJointlyOptimized}.  We also compute (but do not fit with respect to) the squared speed of sound $c_s^2$; this is plotted in Fig.~\ref{FigSpeedOfSoundResults}.  The best fit results are shown in Table \ref{TableOfFitResults} for the different HRG models explored in Refs.~\cite{Albright:2014gva,Albright:2015uua}.

We find that the range of allowable model parameters is constrained by the requirement that the two phases have a coexistence line (where $P_{HRG} = P_{QGP}$) which can correspond to a first-order phase transition line such as that depicted in Fig.~\ref{FigQCDPD}.  This constraint in fact excludes the optimal parameters found in the earlier studies \cite{Albright:2014gva,Albright:2015uua}, meaning that our results do not agree nearly as well with the lattice data as those studies.

\begin{figure*}
\centering
%\minipage{0.5\linewidth}
	\includegraphics[width=0.49\linewidth]{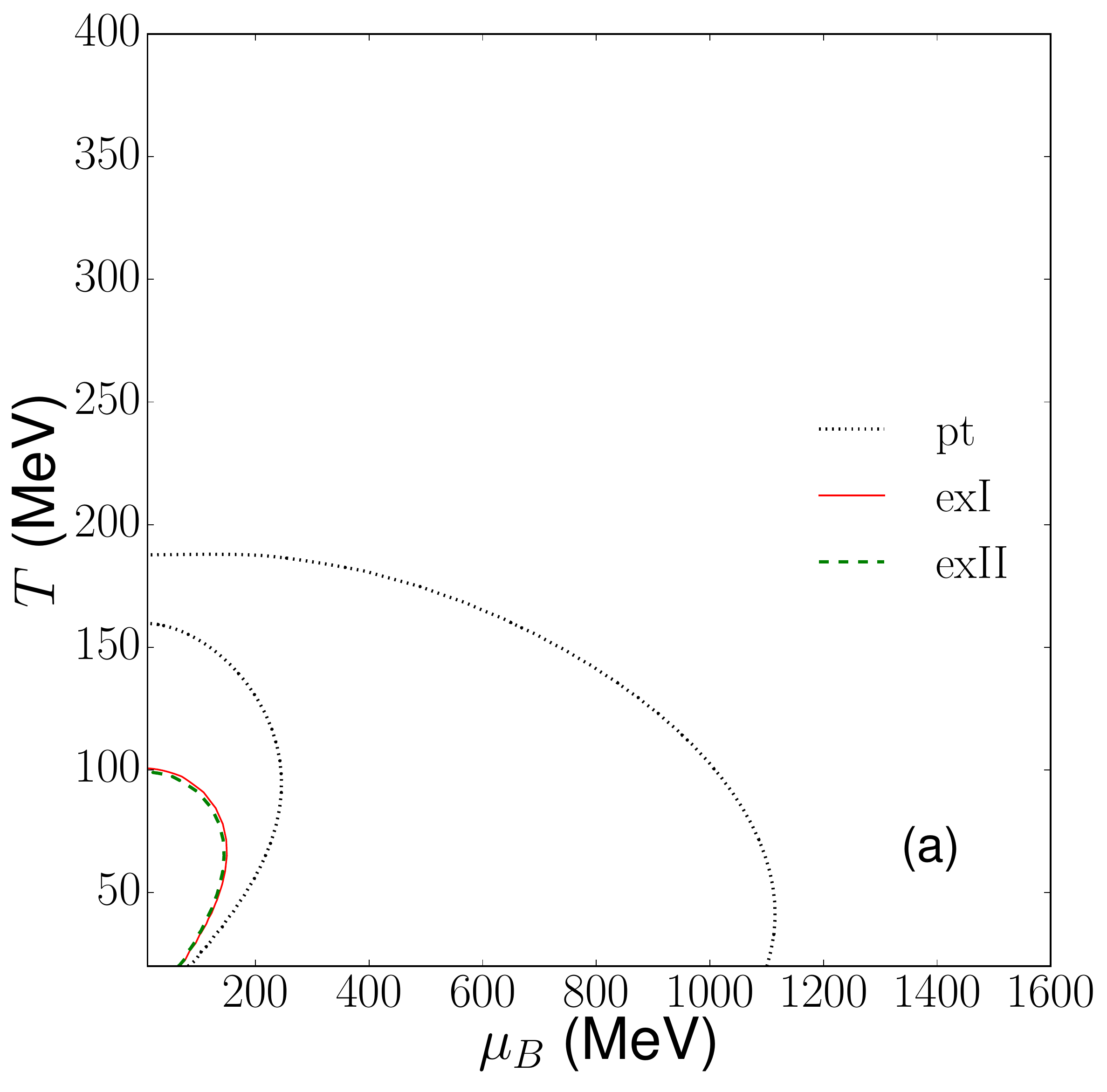}
%\endminipage
\hfill
%\minipage{0.5\linewidth}
	\includegraphics[width=0.49\linewidth]{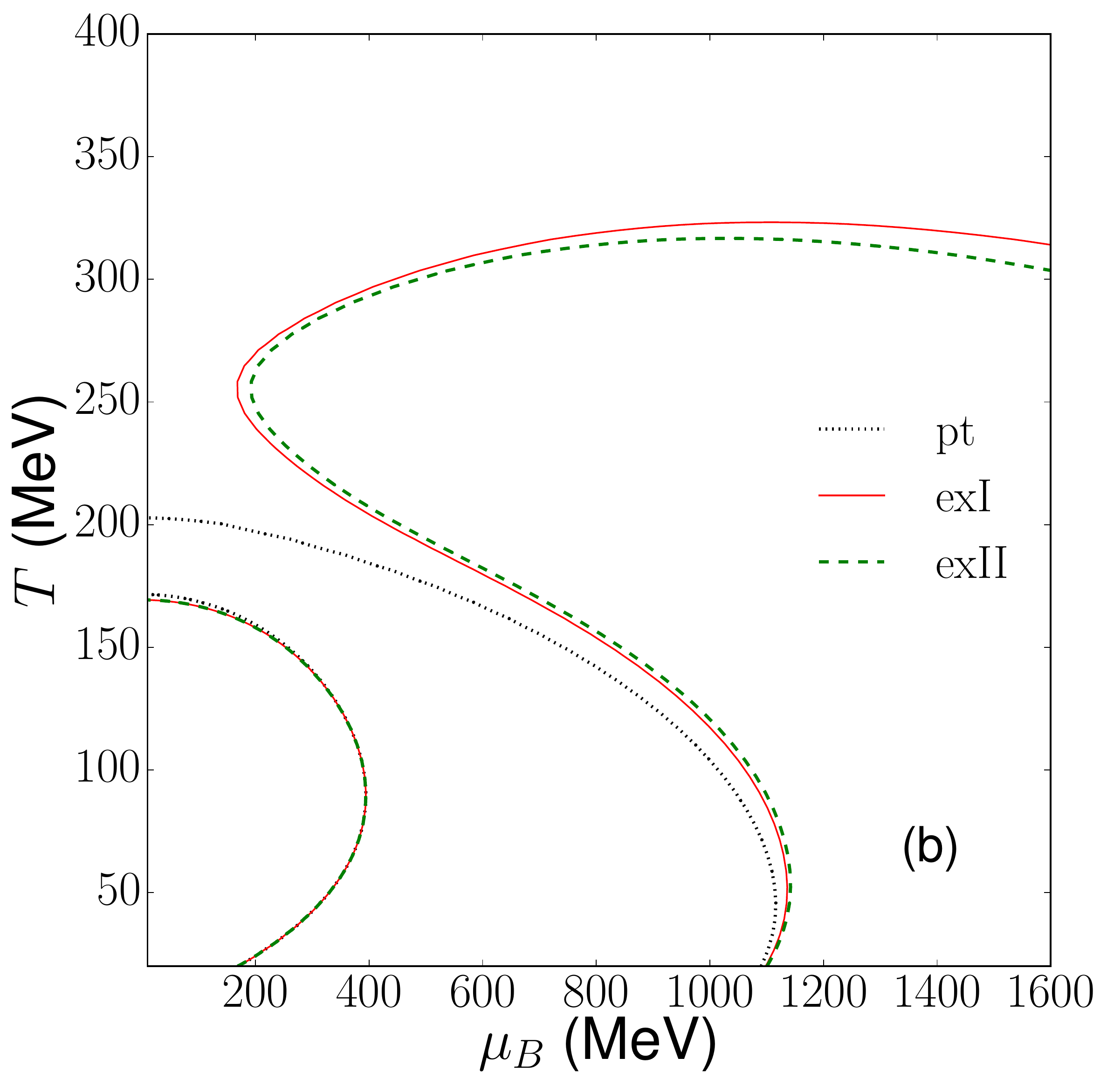}
%\endminipage
	\caption{Crossover curves for various models and parameter combinations.  The lefthand panel corresponds to the optimal parameters obtained in Refs.~\cite{Albright:2014gva,Albright:2015uua}, which clearly do not generally have the proper qualitative behavior in the small-$T$, large-$\mu$ region (with the exception of the pt model, which neglects repulsive interactions).  To obtain the correct behavior restricts our choice of parameters to those which produce curves like the ones shown in the righthand panel.  The latter combinations of model parameters thus entail a maximum possible choice of $T_c$.
	\label{modelCoexistenceLines}}
\end{figure*}

The coexistence curves are illustrated in Fig.~\ref{modelCoexistenceLines} for two different sets of model parameters.  In panel (a), we show the coexistence lines (or lack thereof) for the optimal model parameters used in \cite{Albright:2014gva,Albright:2015uua}.  Clearly, if we are to extend the QGP and HRG phases down to the $\mu$-axis, we will not reproduce a first-order phase transition line as shown in Fig.~\ref{FigQCDPD} for an arbitrary combination of model parameters.  We obtain a more realistic transition line for the combination of model parameters shown in panel (b), at the expense of worsening the overall global fit to the available lattice calculations and requiring a maximum possible choice for the value of $T_c$.  This situation could be improved by using models for $P_{QGP}$ and $P_{HRG}$ which are more reliable in the small-$T$, large-$\mu$ region of the phase diagram and capable of yielding the anticipated shape of the phase transition line near the $\mu$-axis \cite{Kurkela:2016was}.  We plan to explore such improvements in a forthcoming paper \cite{TomChrisJoe}.

%\vspace{15pt}
%%%%%%%%%%%%%%%%%%%%%%%%%%%%%%%%%%%%%%%%%%%%%%%%
\section{Conclusions}
\label{Sec:Conclusions}
%%%%%%%%%%%%%%%%%%%%%%%%%%%%%%%%%%%%%%%%%%%%%%%%
%\vspace*{-2mm}

In this paper, we have constructed a switching function which allows one to obtain an equation of state which switches between separate phases of nuclear matter in a phenomenologically acceptable way.  Switching functions have been constructed for this purpose before \cite{Nonaka:2004pg}, but, to our knowledge, no studies have previously combined a rapid crossover, critical point, and first-order phase transition in a way which explicitly retains infinite differentiability and also enables self-consistent modeling of heavy-ion collisions \textit{via} hydrodynamics.  We have not attempted to incorporate critical indices into our switching function, but we expect that these could be readily included, if desired.  We therefore expect that this model can be put to good use by heavy-ion phenomenologists seeking a relatively simple strategy for implementing phase transitions in a way which is consistent with our current best understanding of the QCD phase diagram.

One particularly valuable aspect of the switching function approach is its generality: we have used particular choices for $P_{QGP}$ and $P_{HRG}$ in this paper, but there are in principle no restrictions on which equations of state one needs to use to describe these two phases.  Our model therefore offers a very flexible approach to modeling transitions between different phases of nuclear matter, and may even be capable of extension to other features of the QCD phase diagram which are not probed by heavy-ion physics, such as the color superconducting phase \cite{Alford:1997zt} or the liquid-gas phase transition at large $\mu$ \cite{Pochodzalla:1995xy}.

\acknowledgments

CP thanks Aleksi Kurkela and Jorge Noronha for useful discussions.  This work was supported by the U.S. Department of Energy (DOE) Grant No. DE-FG02-87ER40328.  The authors also gratefully acknowledge the Minnesota Supercomputing Institute (MSI) at the University of Minnesota for providing resources that contributed to the results presented in this work.


\begin{thebibliography}{99}

\bibitem{Csernai:1992as} 
  L.~P.~Csernai and J.~I.~Kapusta,
  %``Dynamics of the QCD phase transition,''
  Phys.\ Rev.\ Lett.\  {\bf 69}, 737 (1992).

\bibitem{Jacobs:2004qv} 
  P.~Jacobs and X.~N.~Wang,
  %``Matter in extremis: Ultrarelativistic nuclear collisions at RHIC,''
  Prog.\ Part.\ Nucl.\ Phys.\  {\bf 54}, 443 (2005)
  
\bibitem{QM2014}
XXIV International Conference on Ultrarelativistic Nucleus-Nucleus Collisions: Quark Matter 2014, Nucl. Phys. A
{\bf 931}, pp. 1-1266 (November 2014) Edited by P. Braun-Munzinger, B. Friman an
d J. Stachel.

\bibitem{QM2015}
XXV International Conference on Ultrarelativistic Nucleus-Nucleus Collisions: Quark Matter 2015, Nucl. Phys. A
{\bf 956}, pp. 1-974 (December 2016) Edited by Y. Akiba, S. Esumi, K. Fukushima,
 H. Hamagaki, T. Hatsuda, T. Hirano and K. Shigaki.

\bibitem{QM2017}
XXVI International Conference on Ultrarelativistic Nucleus-Nucleus Collisions: Quark Matter 2017, Nucl. Phys. A {\bf 967}, pp. 1-1010 (November 2017) Edited by
U. Heinz, O. Evdokimov and P. Jacobs.

\bibitem{Halasz:1998qr} 
  A.~M.~Halasz, A.~D.~Jackson, R.~E.~Shrock, M.~A.~Stephanov and J.~J.~M.~Verbaarschot,
  %``On the phase diagram of QCD,''
  Phys.\ Rev.\ D {\bf 58}, 096007 (1998)
  
\bibitem{NSAC}
  The Frontiers of Nuclear Science, 2007 NSAC Long Range Plan:
  http://www.sc.doe.gov/np/nsac/docs/Nuclear-Science.Low-Res.pdf

\bibitem{Fukushima:2010bq} 
  K.~Fukushima and T.~Hatsuda,
  %``The phase diagram of dense QCD,''
  Rept.\ Prog.\ Phys.\  {\bf 74}, 014001 (2011)

\bibitem{Csernai:2006zz} 
  L.~P.~Csernai, J.~I.~Kapusta and L.~D.~McLerran,
  %``On the Strongly-Interacting Low-Viscosity Matter Created in Relativistic Nuclear Collisions,''
  Phys.\ Rev.\ Lett.\  {\bf 97}, 152303 (2006)

\bibitem{Bernard:2004je} 
  C.~Bernard {\it et al.} [MILC Collaboration],
  %``QCD thermodynamics with three flavors of improved staggered quarks,''
  Phys.\ Rev.\ D {\bf 71}, 034504 (2005)


\bibitem{Berges:1998rc} 
  J.~Berges and K.~Rajagopal,
  %``Color superconductivity and chiral symmetry restoration at nonzero baryon density and temperature,''
  Nucl.\ Phys.\ B {\bf 538}, 215 (1999)
  
\bibitem{Buballa:2003qv} 
  M.~Buballa,
  %``NJL model analysis of quark matter at large density,''
  Phys.\ Rept.\  {\bf 407}, 205 (2005)


\bibitem{Stephanov:1998dy} 
  M.~A.~Stephanov, K.~Rajagopal and E.~V.~Shuryak,
  %``Signatures of the tricritical point in QCD,''
  Phys.\ Rev.\ Lett.\  {\bf 81}, 4816 (1998)

\bibitem{Fodor:2004nz} 
  Z.~Fodor and S.~D.~Katz,
  %``Critical point of QCD at finite T and mu, lattice results for physical quark masses,''
  JHEP {\bf 0404}, 050 (2004)
  
\bibitem{Stephanov:2004wx} 
  M.~A.~Stephanov,
  %``QCD phase diagram and the critical point,''
  Prog.\ Theor.\ Phys.\ Suppl.\  {\bf 153}, 139 (2004)
  [Int.\ J.\ Mod.\ Phys.\ A {\bf 20}, 4387 (2005)]
  
\bibitem{Iwasaki:2003de} 
  Y.~Iwasaki, K.~Kanaya, S.~Kaya, S.~Sakai and T.~Yoshie,
  %``Phase structure of lattice QCD for general number of flavors,''
  Phys.\ Rev.\ D {\bf 69}, 014507 (2004)


\bibitem{Nonaka:2004pg} 
  C.~Nonaka and M.~Asakawa,
  %``Hydrodynamical evolution near the QCD critical end point,''
  Phys.\ Rev.\ C {\bf 71}, 044904 (2005)


  \iffalse
\bibitem{Kapusta:2010ke} 
  J.~I.~Kapusta,
  %``Equation of State and Phase Fluctuations near the Chiral Critical Point,''
  Phys.\ Rev.\ C {\bf 81}, 055201 (2010)  


\bibitem{Plumberg:2017tvu} 
  C.~Plumberg and J.~I.~Kapusta,
  %``Hydrodynamic fluctuations near a critical endpoint and Hanbury-Brown–Twiss interferometry,''
  Phys.\ Rev.\ C {\bf 95}, no. 4, 044910 (2017)
    
  \fi



\bibitem{Huovinen:2009yb} 
  P.~Huovinen and P.~Petreczky,
  %``QCD Equation of State and Hadron Resonance Gas,''
  Nucl.\ Phys.\ A {\bf 837}, 26 (2010)


\bibitem{deForcrand:2010ys} 
  P.~de Forcrand,
  %``Simulating QCD at finite density,''
  PoS LAT {\bf 2009}, 010 (2009)
  
  


\bibitem{Albright:2014gva} 
  M.~Albright, J.~Kapusta and C.~Young,
  Phys.\ Rev.\ C {\bf 90}, no. 2, 024915 (2014).
  
\bibitem{Borsanyi:2010cj} 
  S.~Borsanyi, G.~Endrodi, Z.~Fodor, A.~Jakovac, S.~D.~Katz, S.~Krieg, C.~Ratti and K.~K.~Szabo,
  %``The QCD equation of state with dynamical quarks,''
  JHEP {\bf 1011}, 077 (2010)
  
\bibitem{Bazavov:2012jq} 
  A.~Bazavov {\it et al.} [HotQCD Collaboration],
  %``Fluctuations and Correlations of net baryon number, electric charge, and strangeness: A comparison of lattice QCD results with the hadron resonance gas model,''
  Phys.\ Rev.\ D {\bf 86}, 034509 (2012)
  
\bibitem{Borsanyi:2012cr} 
  S.~Borsanyi, G.~Endrodi, Z.~Fodor, S.~D.~Katz, S.~Krieg, C.~Ratti and K.~K.~Szabo,
  %``QCD equation of state at nonzero chemical potential: continuum results with physical quark masses at order $mu^2$,''
  JHEP {\bf 1208}, 053 (2012)  
  
\bibitem{Borsanyi:2014ewa} 
  S.~Borsanyi, Z.~Fodor, S.~D.~Katz, S.~Krieg, C.~Ratti and K.~K.~Szabo,
  %``Freeze-out parameters from electric charge and baryon number fluctuations: is there consistency?,''
  Phys.\ Rev.\ Lett.\  {\bf 113}, 052301 (2014)
  
\bibitem{Albright:2015uua} 
  M.~Albright, J.~Kapusta and C.~Young,
  Phys.\ Rev.\ C {\bf 92}, no. 4, 044904 (2015).
  
 \bibitem{Bazavov:2017dus} 
  A.~Bazavov {\it et al.},
  %``The QCD Equation of State to $\mathcal{O}(\mu_B^6)$ from Lattice QCD,''
  Phys.\ Rev.\ D {\bf 95}, no. 5, 054504 (2017)
    
  \bibitem{Kurkela:2016was} 
  A.~Kurkela and A.~Vuorinen,
  %``Cool quark matter,''
  Phys.\ Rev.\ Lett.\  {\bf 117}, no. 4, 042501 (2016)
  
\bibitem{TomChrisJoe}
  T.~Welle, C.~Plumberg, and J.~I.~Kapusta, in progress.
  
\bibitem{Alford:1997zt} 
  M.~G.~Alford, K.~Rajagopal and F.~Wilczek,
  %``QCD at finite baryon density: Nucleon droplets and color superconductivity,''
  Phys.\ Lett.\ B {\bf 422}, 247 (1998)
  
\bibitem{Pochodzalla:1995xy} 
  J.~Pochodzalla {\it et al.},
  %``Probing the nuclear liquid - gas phase transition,''
  Phys.\ Rev.\ Lett.\  {\bf 75}, 1040 (1995).
  
\end{thebibliography}
\end{document}